\documentclass[preprint2]{aastex}

\newcommand{\co}{{\rm $^{12}$CO}}

\newcommand{\hi}{\mbox{\rm \ion{H}{1}}}
\newcommand{\hii}{\mbox{\rm \ion{H}{2}}}
\newcommand{\jone}{($J=1\rightarrow0$)}

\newcommand{\kmpers}{km~s$^{-1}$}

\newcommand{\gpercmsq}{g~cm$^{-2}$}
\newcommand{\gpercmcu}{g~cm$^{-3}$}

\newcommand{\ha}{\mbox{H$\alpha$}}

\newcommand{\mlratiok}{\mbox{$M_\odot/L_{\odot K}$}}

\begin{document}

\title{The Density Profile of the Dark Matter Halo of NGC 4605}
\author{Alberto D. Bolatto, Joshua D. Simon, Adam Leroy, and Leo Blitz}
\affil{Department of Astronomy, University of California at Berkeley, 
\\601 Campbell Hall MC 3411, CA 94720}
\email{bolatto@astro.berkeley.edu}
\begin{abstract}
We have obtained $\sim100$~pc resolution CO and $\sim60$~pc resolution
\ha\ observations of the dwarf spiral galaxy NGC~4605. We use them to
derive a high resolution rotation curve and study the central density
profile of NGC~4605's dark matter halo. We find that these
observations do not agree with the predictions of most high resolution
Cold Dark Matter calculations. We investigate two extreme cases: 1)
NGC~4605 has a maximal exponential disk, that we model using $K$--band
observations and remove to study the structure of its dark matter
halo, and 2) NGC~4605 is dark matter dominated and its disk is
dynamically negligible.  Because the mass--to--light ratio of the
maximal disk is already very low we favor the first solution,
which indicates the halo has one component with a density profile
$\rho \propto r^{-0.65}$ out to $R\sim2.8$ kpc.  In the second
case, the rotation curve requires the presence of two components: a
small ($\sim600$ pc) core surrounded by a much steeper $\rho \propto
r^{-1.1}$ halo.  Removal of intermediate (submaximal) disks does not
ameliorate the discrepancy between the predictions and the
observations.
\end{abstract}

\section{Introduction}

The Cold Dark Matter (CDM) paradigm and its variations (e.g.,
$\Lambda$CDM) have enjoyed remarkable success at explaining the large
scale structure of the Universe. At the scale of individual galaxies,
however, CDM faces several serious challenges
\citep[e.g.,][]{SK00}. One of the key predictions of CDM and
$\Lambda$CDM simulations is that all dark matter halos share a
universal density profile \citep*{NFW96,HJS99}.  The original
calculations predicted a central density profile 
$\rho\sim r^{-1}$, where $r$ is the galactocentric distance.  More
recent, higher resolution simulations have found even steeper
$\rho\sim r^{-1.5}$ central cusps \citep[e.g.,][]{Mea99,Gea00}.

These calculations, however, collide with a number of observational
studies of dark-matter density profiles. Most of the observations find
either central constant density cores
\citep[e.g.,][]{FP94,mGdB98,dBea01,BS01}, or profiles where the
density increases more slowly than $r^{-1}$ toward the center
\citep[e.g.,][]{Kea98}. Other authors conclude that the observational
evidence available to date is inconclusive, largely due to the lack of
high quality high resolution rotation curves \citep{vdBea00}.  Indeed,
using \hi\ interferometers to obtain rotation curve data with the
angular resolution and sensitivity required in order to resolve the
central cusps is a daunting task for all but the nearest galaxies. The
usual alternative, deriving rotation curves from longslit \ha\
spectra, provides spatial information only along one dimension.  This
makes it impossible to separate circular and non circular motions and
to check for position angle errors. Other potential problems with
rotation curves based solely on \ha\ are the systematic effects
introduced by extinction and the local expansion flow of \hii\
regions, which are especially important close to galaxy centers,
precisely where the shape of the rotation curve is most
important. A recent study by \citet*{BOAC01}, however, shows good
agreement between the measured \hi\ and \ha\ kinematics in the inner
parts of their galaxies, indicating that at least in some galaxies
extinction is not a substantial concern. Despite the recent body of
work on the subject, {\em it remains clear that we need more high
quality, high resolution rotation curve measurements to advance
towards solving the galactic scale dark matter puzzle.}

Since both \hi\ and \ha\ rotation curve studies are problematic, we
have opted for a third, complementary technique: we use CO
interferometry to measure the inner rotation curves.  Millimeter wave
interferometry avoids the problems inherent to the first two
techniques. First, it is much more sensitive to surface brightness
than \hi\ interferometry, allowing us to routinely obtain sensitive
data with 5\arcsec\ angular resolution or better while the typical
resolution of most \hi\ studies is $\sim30\arcsec$.  Second, not only
is CO interferometry unaffected by extinction, but it works best
near galaxy centers where most molecular clouds are typically
found. Third, like \hi\ interferometry but unlike long slit \ha\
spectroscopy, it provides complete spectral imaging information
allowing us to obtain reliable rotation curves.

The relative sensitivity advantage of CO interferometry stems from the
fact that the \co\ \jone\ rotational transition at 2.6~mm has a
wavelength $\sim80$ times smaller than the 21~cm spin flip \hi\
transition. This allows the CO antennas to be placed $\sim80$ times
closer together than the \hi\ antennas for the same angular
resolution. Because the sensitivity of aperture synthesis to surface
brightness is proportional to the fraction of the filled aperture,
compactness confers a considerable sensitivity advantage to
millimeter wave interferometers. For example, the ten element
Berkeley--Illinois--Maryland Association millimeter array
\citep[BIMA;][]{Wea96} has a 250 times larger filled aperture fraction
than the VLA.

The difficulty in using CO interferometry for dark matter studies,
however, is finding suitable target galaxies. Late type dwarf galaxies
are ideal for such studies, because their density profiles are
dominated by dark matter in their inner regions
\citep*{PS90,SAP91}. Although little is known about the molecular gas
content of most dwarf galaxies, with a few notable exceptions they are
not strong molecular sources. We have obtained CO interferometer maps
of several nearby dwarf galaxies with single--dish detections, a
subset of which we selected for rotation curve studies.  To further
expand the set of suitable sources we are carrying out a systematic
survey of nearby dwarf galaxies in CO. Thus, we expect that the
results presented here are only a first glimpse of the final study.
We present our observations in \S\ref{observations}, perform a
rotation curve study in \S\ref{discussion}, and summarize our results
in \S\ref{conclusions}.

\section{Observations and Data Processing}
\label{observations}

The source we discuss in this work, NGC 4605, is an SBc dwarf galaxy.
Independent photometric estimates place this galaxy $\sim4.6$ Mpc
\citep{KT94} and $4.26\pm0.64$ Mpc away (M. Pierce, priv. comm.),
yielding a physical scale of $\sim20$ pc/\arcsec. The physical
resolution of our CO observations (and of our rotation curve) is
therefore $\sim100$ pc. The extinction corrected absolute B magnitude
of NGC~4605 is $M_B=-18.10$ and its size is $R_{25}\sim3\arcmin$
($\sim3.6$ kpc), very similar to the magnitude and size of the Large
Magellanic Cloud (LMC). The line-of-sight inclination of NGC 4605 is
$i\approx69\deg$, and \hi\ observations find an inclination corrected
linewidth at 20\% intensity $W_{20}\approx198$ \kmpers\
\citep{Bea85}. This indicates a maximum rotational velocity of
$\sim100$ \kmpers\ (again, similar to that of the LMC) that along with
its size and luminosity definitively identifies this galaxy as a
dwarf. Carbon monoxide millimeter emission was detected in NGC 4605 by
\citet{Yea95} using the FCRAO 14 m radio telescope. Our first
interferometer maps revealed CO emission throughout the disk of NGC
4605, which became the obvious first choice for a rotation curve
study.

\begin{figure}
\plotone{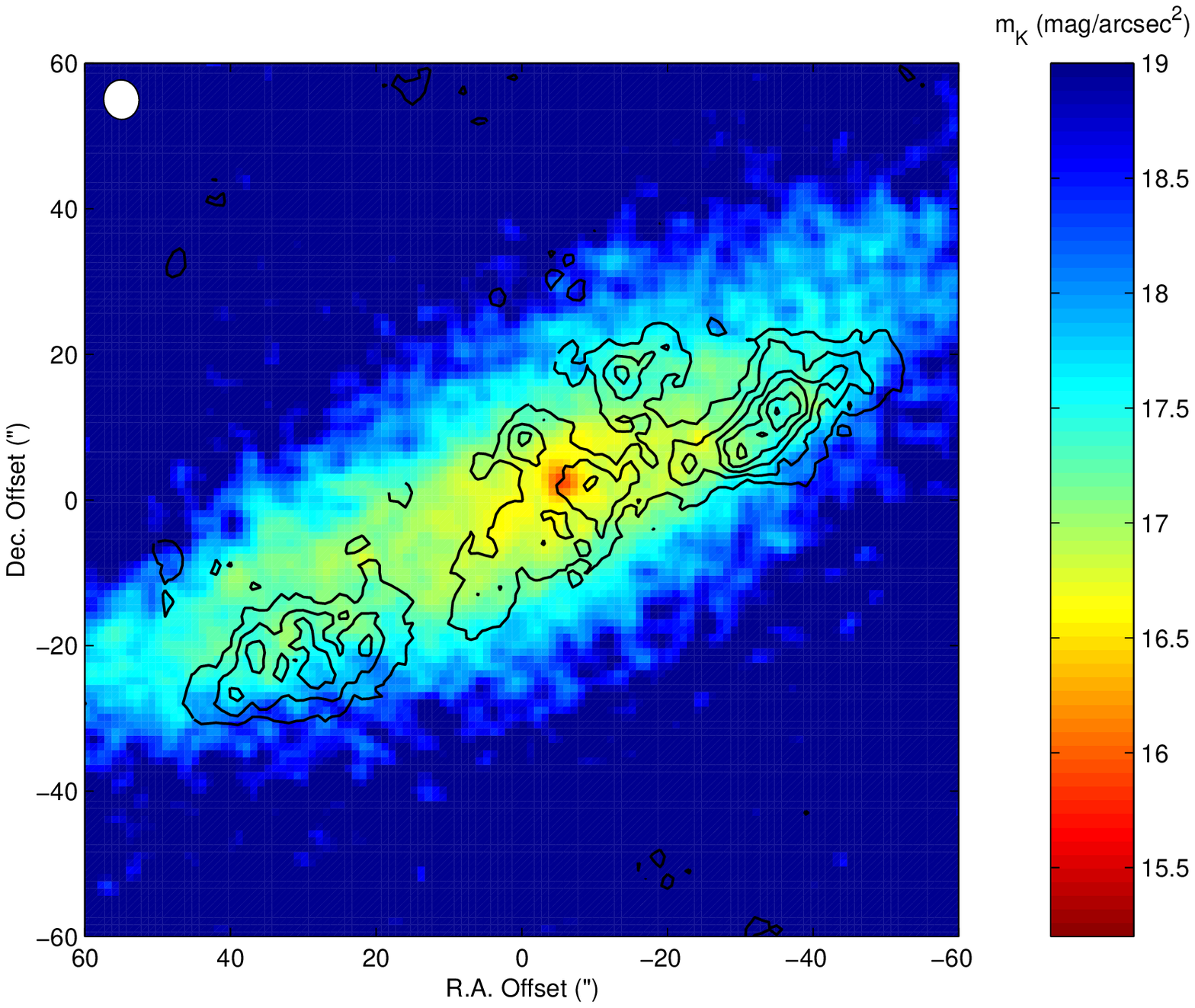}
\caption{2MASS $K$--band image of NGC~4605 with integrated intensity
CO contours overlaid. The RMS of the integrated intensity data is
$\sim130$ mJy beam$^{-1}$ \kmpers\ ($\sim450$ mK \kmpers). The contours
start at $3\sigma$ and increase in steps of $4\sigma$ to 3 Jy
beam$^{-1}$ \kmpers. The offsets are with respect to NGC~4605's
center, at $\alpha_{J2000}=12^{\rm h}40^{\rm m}00\fs0$,
$\delta_{J2000}=61\arcdeg36\arcmin31\arcsec$. The synthesized beam is
shown in the upper left corner.
\label{intmap}}
\end{figure}

We observed the \co\ \jone\ rotational transition in NGC 4605 using
three configurations of the BIMA array (B, C, and D) between June 2000
and March 2001. We mapped one BIMA primary beam, subtending a field
with a half power diameter of $\sim100\arcsec$, sufficient to
encompass the CO emission. The spectrometer was configured with 2
\kmpers\ wide channels and a bandpass of 260 \kmpers. The individual
tracks were calibrated, combined, imaged, and deconvolved using the
CLEAN algorithm within the MIRIAD astronomical package. The resulting
naturally weighted map has a beam size of $4.8\arcsec\times5.4\arcsec$
($\sim99\times111$ pc) with $P.A.\approx6\deg$ (Fig. \ref{intmap}).
The individual planes of the datacube have an RMS of $\sim24$ mJy
beam$^{-1}$ ($\sim85$ mK) at a velocity resolution of 3 \kmpers.

\begin{figure}
\plotone{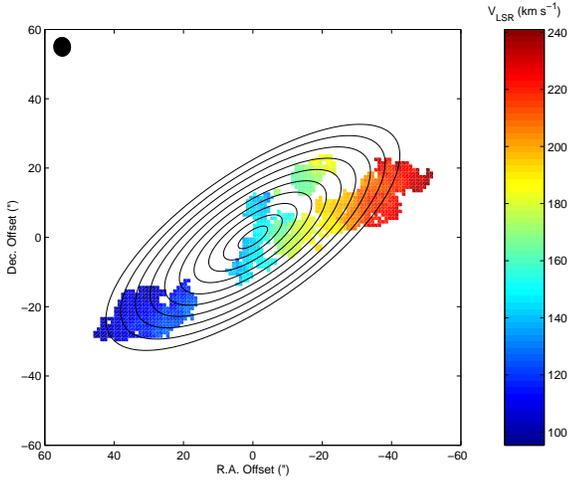}
\caption{Velocity field obtained from the CO datacube. The typical
$1\sigma$ error in the velocities is $\sim2 - 3$ \kmpers. The ellipses
correspond to the projection of the 5\arcsec\ wide rings used to
determine the rotation curve.
\label{velmap}}
\end{figure}

\begin{figure}
\plotone{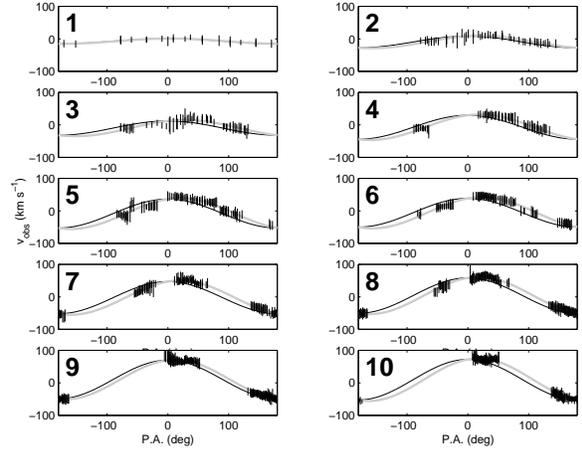}
\caption{Individual fits to the CO rings, velocity plotted against
angle. The number in the upper left corner indicates the ring number,
increasing outwards. The vertical lines are the individual data points
with their errors, including the square root of oversampling
factor. The gray thick line is the overall fit to $v_{obs}$, while the
black thin line is the $v_c$ component, as described in the main text.
\label{annuli}}
\end{figure}

\begin{figure}
\plotone{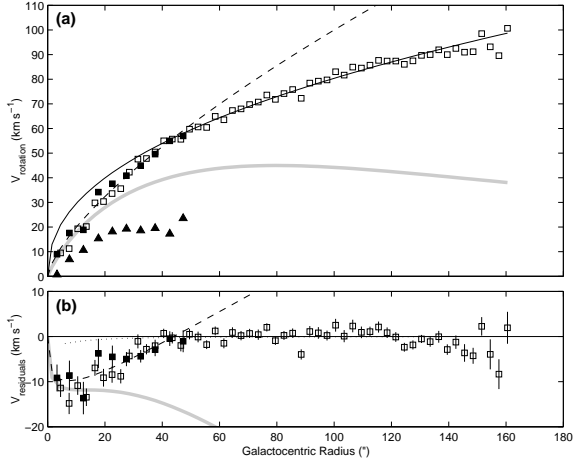}
\caption{Combined CO and \ha\ rotation curve of NGC~4605 with no disk
removed, in linear--linear scale. Plot (a) shows the individual
rotation measurements for CO (filled squares) and \ha\ (open
squares). It also shows the $v_r$ component obtained from the CO (filled
triangles). The CO is sampled every 5\arcsec, the \ha\ every 3\arcsec,
and both have been corrected for the effects of beam smearing.  The
solid line is the power law fit to the outer halo \ha\ data in the
$r\in[40\arcsec,140\arcsec]$ region, $\rho\propto r^{-1.1}$. The dashed
line is the power law fit to the CO rotation, yielding $\rho\propto
r^{-0.6}$. The gray thick line is the maximal disk obtained from the
$K$--band observations. Plot (b) shows the residuals in the data and
the fits, after removing the outer halo fit.  The dotted line is the
effect of beam smearing, as it was removed from the \ha\
observations. The correction applied to the CO data is very similar.
The error bars are internal, statistical errors obtained from the fits.
\label{rotcurve}}
\end{figure}

The datacube was used to produce a first moment map of the CO emission
(Fig. \ref{velmap}). The map was rotated to align the optical major
axis of the galaxy with one of the coordinate axes ($P.A.=125\deg$),
and deprojected using the optical center and inclination angle. The
velocity data were corrected by the sine of inclination angle, and
sorted in concentric rings of width $5\arcsec$. To each ring we fit a
model $v_{obs} = v_{t} + v_c \cos\theta + v_r \sin\theta$,
representing the projected effects of translation ($v_{t}$), circular
($v_c$), and radial ($v_r$) velocities. The errors in the velocity
map, multiplied by the square root of the beam oversampling factor,
are used to weight each point in the fit (Fig. \ref{annuli}). The
error bars for the fits (which range from 1.4 to 3.6 \kmpers) are
computed using the diagonal of the covariance matrix. We find a small
$v_r$ component, increasing outwards to $\sim25\arcsec$ and then
leveling off (Fig. \ref{rotcurve}). This component is never larger
than half $v_c$, and adding it in quadrature to $v_c$ does not
significantly change the density profile exponents discussed in this
paper. The presence of this component may indicate that the kinematics
inside $\sim25\arcsec$ have a position angle slightly different from
the optical $P.A.$ (closer to $P.A.\sim95\deg$).

To extend the CO rotation curve beyond 50\arcsec\ ($\sim1$ kpc), we
observed NGC~4605 with the 0.6~m Coud\'e Auxiliary Telescope at Lick
Observatory on the night of 23 April 2001.  We used the Hamilton
Echelle Spectrometer with a 75\AA\ wide filter centered on \ha\
($\lambda=6562.8$\AA) to block adjacent spectral orders.  The slit was
640 $\mu$m (6\arcsec) wide and $\sim8\arcmin$ long, oriented along
NGC~4605's major axis at $P.A.=125\deg$ using the facility's focal
plane image rotator. Our velocity resolution was $\sim6$ \kmpers, with
a spatial resolution of $\sim3\arcsec$.  This spectral resolution is
much higher than that of most published rotation curve studies, and it
allows us to very accurately measure the kinematics of the inner disk
of NGC~4605. We integrated for 4.5 hours divided in 30 minute
exposures using manual guiding.

We reduced the \ha\ data using IRAF. We removed a bias frame from the
individual exposures, and combined them using the {\sc imcombine}
routine with its built in cosmic ray rejection algorithm.  Because no
continuum was detected in the individual images, we registered them
using offsets obtained by cross correlating the spectra.  For
wavelength calibration we used exposures of a ThAr reference lamp, and
observations of a radial velocity standard star. To extract the
rotation velocities we fit a gaussian to each row of the spectrum.  We
folded the results into a rotation curve by defining the center of the
galaxy to be the point at which the two sides of the rotation curve
lined up best. The typical errors in the individual points of the
folded rotation curve were less than 2 \kmpers. Comparison of the CO
and \ha\ rotation data reveals that both tracers are in good agreement
over the common region (Fig. \ref{rotcurve}). Thus \ha\ appears to be
a good tracer of the kinematics of the inner disk in NGC~4605. 

We developed a simulation to estimate how our finite angular
resolution affects both the CO and \ha\ rotation curves (beam
smearing). Because calculating the effect of the beam smearing
requires assuming an underlying velocity field, we performed these
computations in an iterative, self consistent manner. The net effect
is very small, amounting to a velocity loss of $\sim 2$ \kmpers\ for
the innermost points and rapidly decreasing outwards. Although small,
this systematic velocity loss is not entirely negligible in the inner
30\arcsec.  We removed the effect of beam smearing from the data
before performing the fits.

We used 2MASS $K_s$ data to derive the distribution of mass in the
disk of NGC~4605. The advantage of using $K$--band observations to
model the disk mass over more typical $B$, $V$, or $I$ data cannot be
overemphasized: at $K$--band the extinction is a factor of $\sim10$
down from $V$, and a factor of $\sim5$ down from $I$ \citep{RL85}.  We
retrieved the archival 2MASS atlas image for NGC~4605 and extracted
the surface brightness profile using the photometric calibration
provided in the FITS header. We then fit an exponential disk to the
surface brightness profile using data with galactocentric radii
$r\in[10\arcsec,60\arcsec]$, finding $\mu_K\approx16.35+0.029\,r$
where $\mu_K$ is the apparent $K$--band magnitude. NGC~4605 does not
have a significant bulge, and the surface brightness distribution is
well fit by an exponential law over this range.

\section{Results and Discussion}
\label{discussion}

The universal dark matter halo density law proposed by \citet*{NFW95}
follows the form

\begin{equation}
\frac{\rho(r)}{\rho_{crit}} = \frac {\delta_c}{(r/r_s)(1+r/r_s)^2} \ ,
\end{equation}

\noindent where $\rho_{crit}=3H^2/8\pi G\sim1\times10^{-29}$
\gpercmcu\ is the critical density, $\delta_c$ is the halo
overdensity, and $r_s$ is the characteristic radius (simulations
suggest $r_s$ of several kpc). At galactocentric distances $r\ll r_s$
the halo has an approximate density profile
$\rho\approx\rho_{crit}\,\delta_c\,r_s\,\,r^{-1}$. In general, 
a density profile $\rho\propto r^{-\alpha}$ implies

\begin{equation}
v_c\propto r^{(2-\alpha)/2} \ .
\label{alphaeq}
\end{equation}

Most dark matter density profile studies assume that the halo is the
dominant mass component, and do not attempt to model and remove the
contribution from the galactic disk to the rotation, which is taken to
be negligible. We call this approach the ``minimal disk'' hypothesis,
since it implies that the mass--to--light ($M/L$) ratio of the
material in the disk is $M/L\sim0$. The opposite limit is given by the
``maximal disk''; the maximum exponential disk that can be subtracted
from the measured rotation curve without creating negative
residual velocities. Fitting a maximal disk to the rotation curve yields a
maximum mass--to--light ratio for the material in the disk
(predominantly stars and gas).

To find the upper limit to the $M/L$ ratio in NGC~4605 we fit a thin
maximal exponential disk using our rotation curve and the 2MASS
$K$--band data. The fit is constrained by the innermost points of the
rotation measurements. This fit yields $M/L_K=0.25\pm0.03$
\mlratiok\ (we assume $M_{\odot K}=3.34$), implying a disk mass
surface density of stars plus gas $\mu=0.053\exp(-r/37\farcs1)$
\gpercmsq. The extinction corrected color, $B-R=0.67$ (M. Pierce,
priv. comm.), predicts a similar mass--to--light ratio $M/L_K\sim0.34$
\citep{BdJ01}.  This value is very close to the lower limit of the
reasonable range of $M/L_K$, possibly corresponding to a young
starburst population according to the Starburst99 models
\citep{Lea99}.  This idea is further supported by the observation
that, in dwarf galaxies, bright CO emission appears to be accompanied
by the intense star formation activity signaling a young
starburst. The fact that the upper limit of $M/L$ is so low suggests
that the disk of NGC~4605 is quite close to maximal, as submaximal
disks will imply even lower values of the mass--to--light ratio. Using
the single dish CO and \hi\ data and allowing for Magellanic
metallicities we estimate the gas is $\sim20\%$ of the total mass of
the maximal disk. The effect of including the gas would be to lower
the stellar mass--to--light ratio to $M/L_K\sim0.20$
\mlratiok. Because this correction is somewhat uncertain we ignore it
in the following discussion, but note that its only effect is to make
the $M/L$ ratio even lower.

\begin{figure}
\plotone{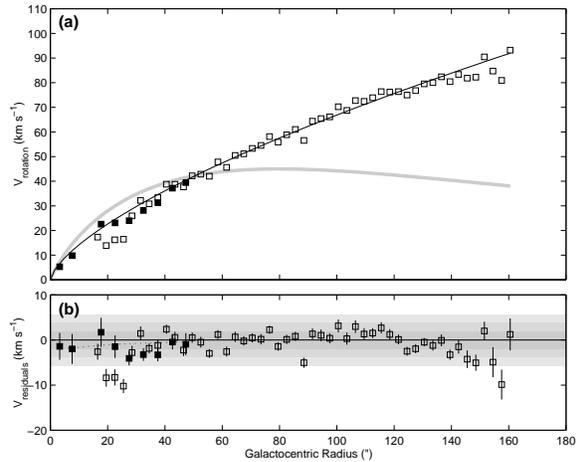}
\caption{Combined CO and \ha\ rotation curve of NGC~4605 after
removing the maximal disk. The key to symbols and lines is the same as
in Fig. \protect\ref{rotcurve}. The few data points missing had $v_c <
v_{disk}$, yielding imaginary $v_{halo}$. Plot (a) shows the new outer
halo fit for $r\in[40\arcsec,140\arcsec]$, $\rho\propto r^{-0.65}$, and
the fact that the maximal disk is dynamically dominant inside
$r\sim50\arcsec$. Plot (b) shows the residuals after removing the
outer halo fit. The gray regions indicate $1\sigma$, $2\sigma$, and
$3\sigma$ departures from the fit. There is no evidence for a break in
the power law (i.e., a core) after the disk is removed.
\label{maxdisk}} 
\end{figure}

The rotation due to an exponential disk is straightforward to compute
\citep{Fr70}, and to remove it from the measured rotation we use
$v^2_{c}=v^2_{halo}+v^2_{disk}$, where all velocities are circular
velocities \citep{BT87}. Figure \ref{maxdisk} shows the result of
removing the maximal disk contribution from the overall rotation. To
the residual rotation, due to the halo, we fit a $v_{halo}\propto
r^{\beta}$ power law in the region $r\in[40\arcsec,140\arcsec]$ and
find the corresponding $\alpha$ using Eq. \ref{alphaeq}. The best fit
power law is $v_c\sim3.0\,r^{\,0.67}$ \kmpers, implying
$\rho\sim4.9\times10^{-24}\,R^{-0.65}$ \gpercmcu\ (where $R$ is the
galactocentric radius in kpc). This value of $\alpha$ is significantly
lower than the calculations by \citet{NFW96} or \citet{Mea99}
indicate. Note, however, that it agrees well with the critical
solution recently found by \citet{TN01} with $\alpha\sim0.75$. Thus,
{\em removing the maximal disk from the rotation curve of NGC~4605
results in a shallow halo, with a density exponent
$\alpha=0.65\pm0.02$}. The error bars including the uncertainty
in the $M/L$ ratio are $\pm0.07$.

It is interesting to investigate whether the halo rotation of NGC~4605
shows any hint of a constant density core, such as those proposed for
other galaxies \citep[e.g.,][]{BS01}. The region inward of 30\arcsec\
in Fig. \ref{maxdisk}b does not show a significant departure from the
outer halo fit, especially after correcting the innermost points for
the effects of beam smearing. While most velocity residuals inside
30\arcsec\ are negative, the overall deficit is less than 2 \kmpers.
Because the maximal disk is dynamically dominant inside $r=40\arcsec$
any departures from a perfect exponential disk, or from a constant
mass--to--light ratio, can account for it.  Among other problems that
could introduce similar systematic errors near the centers of galaxies
are: 1) a displacement of the \ha\ slit by a few arcseconds
perpendicular to the major axis of the galaxy, 2) a disagreement of a
few arcseconds between the optical and dynamical centers, or 3) the
presence of a weak galactic bar with the consequent non circular
orbits. For these reasons, it is difficult to completely exclude a
constant density core using these data: in the presence of a
dynamically dominant disk the rotational signature of such core is a
deficit of a few \kmpers\ throughout this region. Thus, although we
find no evidence for a constant density core, we cannot discard it
completely.

What happens if despite the evidence for a maximal disk in NGC~4605 we
assume the minimal disk hypothesis? It is easy to see that the
observed rotation cannot be fit well by a single power law
(Fig. \ref{rotcurve}).  This suggests that there are at least two
distinct mass components that are contributing to the potential.
Indeed, a possible decomposition of the rotation curve of NGC~4605
indicates a slowly varying density profile inside 30\arcsec\ (i.e., a
core with $v_c\sim3.0\,r^{\,0.8}$ \kmpers, implying
$\rho\sim1.5\times10^{-23}\,R^{-0.4}$ \gpercmcu), and a rapidly
decreasing density outwards of 40\arcsec\ ($v_c\sim10.4\,r^{\,0.45}$
\kmpers, yielding $\rho\sim 8\times10^{-24}\,R^{-1.1}$
\gpercmcu). {\em In the absence of a disk the observed rotation
requires a two component halo: a core with $\alpha=0.4\pm0.2$
inside $r\sim30\arcsec$ and an outer halo with
$\alpha=1.1\pm0.02$.}

Both of these limits are in disagreement with most results of
collisionless cold dark matter simulations. The removal of a
submaximal disk does not eliminate the discrepancy. Such a solution
has a combination of the problems of each limit, requiring both a
shallow outer halo and an inner core. It should be noted that the
goodness of fit of the outer halo by a power law (or any other
mathematical model used to describe it) cannot be reliably employed to
obtain a ``best fit'' mass--to--light ratio and select a
solution. Indeed, we find that for $M/L$ in the range $M/L\in[0,0.25]$
\mlratiok, the RMS of the residuals of the power law fit to the
rotation curve only changes from 1.5 (Fig. \ref{rotcurve}b) to 1.9
\kmpers\ (Fig. \ref{maxdisk}b), showing that the RMS is very
insensitive to how much of the stellar disk is removed.  We have
argued that, because we find a low mass--to--light ratio, NGC~4605's
disk is close to maximal. This solution has the virtue of requiring
only one halo component. Regardless of the precise $M/L$ ratio of the
disk of NGC~4605, the inescapable conclusion is that the shape of the
halo of NGC~4605 poses a challenge to the CDM simulations.

\section{Summary and Conclusions}
\label{conclusions}

We have measured the kinematics of the dwarf galaxy NGC~4605 with high
spectral ($\sim2$ \kmpers\ for the CO, $\sim6$ \kmpers\ for the \ha)
and spatial ($\sim100$ pc) resolution. Our study of the combined CO
and \ha\ rotation curve leads to a robust conclusion: the standard
collisionless CDM simulations do not accurately predict its shape.  We
have discussed two extreme cases: 1) the measured rotation is due to a
combination of the halo and an exponential maximal disk, and 2) the
rotation is due to solely the halo (i.e., the minimal disk
hypothesis).  Because the $K$--band mass--to--light ratio we find for
the first case is very low ($M/L_K\approx0.25$ \mlratiok), we believe
that the disk of NGC~4605 is close to maximal and that the first
solution is the correct one. This solution requires only one halo
component within $R\sim2.8$ kpc. The resulting $\rho\sim r^{-\alpha}$
halo density profile has $\alpha=0.65\pm0.07$. This exponent is
substantially lower than the predictions of most high resolution CDM
calculations, although it agrees much better with the critical
solution found by \citet{TN01}.

If we disregard the evidence favoring a maximal disk and adopt the
minimal disk case, we find that we need a soft $\rho\sim r^{-0.4}$
core to explain the rotation of the innermost $600$ pc. Note that
while the precise value of $\alpha$ for this core is rather uncertain,
the fact that a break in $\alpha$ is required is not. Furthermore,
the density profile of the halo in the $0.8$ to $2.8$ kpc region has
$\alpha=1.1\pm0.02$. This is still noticeably different from the
results of some of the recent high resolution simulations which find
$\alpha\approx1.5$ \citep{Mea99,Gea00}, although it agrees with other
calculations that find $\alpha\approx1.1$ \citep{JS00}. None of these
calculations, however, predict the core. The removal of a
submaximal disk has a combination of the problems of these two
extreme cases.

Notwithstanding which solution is the best, both of these extremes
(and all of the intermediate cases) are discrepant with most of the
available predictions of standard CDM. We have to note that the
strength of these conclusions stems from the fact that both the CO and
the \ha\ data, independently, reveal the same kinematics for the inner
$50\arcsec$ of NGC~4605. A similar analysis, relying only on one or
the other rotation curves, would be impaired by the possible
systematics of the individual datasets.

\acknowledgements We wish to thank the anonymous referee for a speedy
and thorough review of this paper.  We wish to acknowledge the help of
James Graham, and the support of the RAL and the BIMA team, especially
R. Forster, R. Gruendl, R. Plambeck, and E. Sutton.  This research was
supported by NSF grant AST-9981308, and made extensive use of the
NASA/IPAC Extragalactic Database (NED), the Los Alamos National
Laboratory astrophysics preprint database, and NASA's Astrophysics
Data System Abstract Service (ADS).

\end{document}